\documentclass[12pt,a4paper,final,showpacs,amsmath,amssymb,aps,notitlepage]{iopart}
\usepackage{bbm}
\usepackage{braket}
\usepackage{amsthm}
\usepackage{mathrsfs}
\usepackage{txfonts}
\usepackage{bm}
\usepackage[dvipdfm]{graphicx}
\usepackage{color}
\usepackage{cite}
\usepackage{subfigure}
\usepackage{wrapfig}

\begin{document}

\title{Security of differential quadrature phase shift quantum key distribution}
\author{
Shun Kawakami, Toshihiko Sasaki and Masato Koashi}

\address{
Photon Science Center, Graduate School of Engineering, The University of Tokyo, 
2-11-16 Yayoi, Bunkyo-ku, Tokyo 113-8656, Japan}

\ead{kawakami@qi.t.u-tokyo.ac.jp}

\begin{abstract}
One of the simplest methods for implementing quantum key distribution 
over fiber-optic communication is the Bennett-Brassard 1984 protocol 
with phase encoding (PE-BB84 protocol), in which the sender uses phase 
modulation over double pulses from a laser and the receiver uses a 
passive delayed interferometer. 
Using essentially the same setup and by 
regarding a train of many pulses as a single block, one can carry out 
the so-called differential quadrature phase shift (DQPS) protocol, which 
is a variant of differential phase shift (DPS) protocols. 
Here we prove the security of the DQPS protocol based on an adaptation 
of proof techniques for the BB84 protocol, which inherits the advantages 
arising from the simplicity of the protocol, such as accommodating the 
use of threshold detectors and simple off-line calibration methods for 
the light source. We show that the secure key rate of the DQPS protocol 
in the proof is eight thirds as high as the rate of the PE-BB84 protocol.


\end{abstract}

\maketitle

\section{Introduction}
Quantum key distribution (QKD) aims to realize private communication securely from an adversary with 
unlimited power, using the properties of quantum mechanics.  
Among many applications of quantum information, the QKD is relatively matured 
from both theoretical and experimental aspects, with a lot of practical demonstrations already conducted 
\cite{2009PeevSECOQC,2011Tokyo}. 
From a practical viewpoint, it is desired that a QKD protocol is implemented with a conventional laser as 
its light source, and with simple hardware for encoding, decoding and detection. 
The simplicity is desired not only because of a lower cost and a higher clock rate, 
but also because complicated systems and procedures tend to impose severe restrictions on the model of 
the sender's and the receiver's apparatus, 
and to suffer from inefficiency in short-time communications due to a large overhead 
involved in statistical estimations.  
The BB84 protocol with phase encoding (Phase-encoding BB84, PE-BB84 henceforth)
\cite{1984Bennett,1995MarandPhase-Encoding}, 
which uses four relative phases $\{0,\frac{\pi}{2},\pi,\frac{3\pi}{2}\}$ between two neighboring pulses, 
is one of the simplest QKD implementations suited for communication over optical fibers. 
In the PE-BB84 protocol, the sender only needs a phase modulator for encoding, 
while the receiver only needs a passive Mach-Zehnder interferometer with two detectors. 
With its established security\cite{1996Mayers,2000Shor,2004GLLP,2009Koashi}, 
a number of demonstrations have so far been reported \cite{2004Gobby,2007RosenbergPEBB84,2013Shields}. 

For long-distance communication, the laser-based BB84 protocol suffers from photon-number splitting (PNS) 
attacks \cite{2000Brassard}. 
It is often used with decoy-state method \cite{2005Wang,2005Decoy} to add protection against such attacks, 
but the decoy-state method sacrifices the simplicity of the PE-BB84 protocol and benefits associated with it, 
requiring additional devices as well as severer physical assumptions on the light source. 
In contrast, the differential-phase-shift (DPS) QKD \cite{2002Inoue} was proposed 
to achieve protection from the PNS attacks 
while retaining (or even improving) the simplicity of the PE-BB84 protocol. 
The DPS protocol uses two relative phases $\{0,\pi\}$ between every neighboring pair of pulses belonging to 
a long train of pulses, and demonstration with a high clock rate was conducted\cite{2007Takesue}. 
Although its security was proved and the robustness against PNS attacks was verified, 
so far the key generation rate is much smaller than the decoy-BB84 protocol\cite{Tamaki2012}. 
A new approach to improve the key rate was also proposed \cite{Sasaki2014} 
and demonstrated \cite{2015Guan,Takesue2015a,Li2015,Wang2015} recently, 
but it requires an additional element 
in the receiver's apparatus to measure relative phases between pulses at different intervals.

In this article, we seek after the benefit of the DPS QKD in a different direction, 
namely, for short-distance communication in which a PNS attack does not impose a severe problem. 
We provide a security proof of a variant of the DPS QKD called differential quadrature phase shift 
(DQPS) protocol\cite{2009Inoue}, and establish that it has a definite advantage over the PE-BB84 protocol. 
The DQPS protocol can be implemented with essentially the same hardware as the PE-BB84 protocol, 
but our security proof shows that its key generation rate is 8/3 as high as that of the PE-BB84 protocol. 
The benefit from the simplicity of PE-BB84 protocol is not sacrificed because 
the requirement for the properties of the light source and the detection apparatus is shown to be kept to 
minimum as in the PE-BB84 protocol. 

This paper is organized as follows. In Sec.~2, we describe details of the DQPS protocol 
and assumptions on the light source and the detection apparatus. 
Sec.~3 gives the security proof of the protocol, and shows an explicit formula for the key rate. 
Based on the formula, numerical results for the secure key rate is shown in Sec.~4. 
Finally, Sec.~5 deals with discussions including an analytical expression for the scaling of the optimal key 
rate.

\section{Protocol}\label{protocol}
\begin{figure}[htbp]
  \begin{center}
  \raisebox{5mm}
  {\includegraphics[width=150mm]{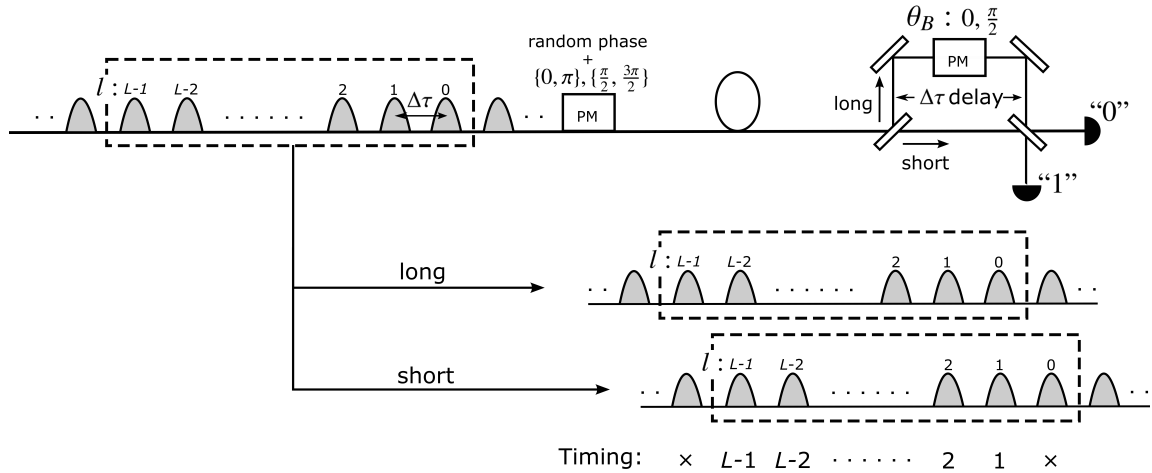}}
  \end{center}
  \begin{center}
  \caption{Setup for the $L$-pulse DQPS protocol. 
  At Alice's site, pulse trains are generated by a laser 
  followed by phase randomization as well as phase modulation (PM) 
  with $\{0,\pi\}$ or $\{\frac{\pi}{2},\frac{3\pi}{2}\}$ according to her random bits and basis choice. 
  At Bob's site, 
  each pulse train is fed to a delayed Mach-Zehnder interferometer with phase shift 
   0 or $\frac{\pi}{2}$ according to his basis choice.  
   The trains leaving the interferometer are measured by two photon detectors corresponding to bit 
   values ``0'' and ``1''.  
   Valid timings of detection are labeled by integers $1,2,..,L-1$, 
   according to the index of the pulse from the short arm of 
   the interferometer. Detection from  
   interference between pulses from different blocks is regarded as invalid and ignored. 
    \label{setupfig}
    }
   \end{center}
\end{figure}
Here we introduce a DQPS protocol considered in this work, which is slightly modified from the one 
\cite{2009Inoue} proposed by Inoue and Iwai (See  Fig.~\ref{setupfig}). 
The protocol uses  two bases, data basis for generating the final key  and check basis 
for monitoring the leak of information. 
In the data and check bases, relative phases between adjacent pulses are modulated by $\{0,\pi\}$ and 
$\{\frac{\pi}{2},\frac{3 \pi}{2}\}$, respectively.
The protocol regards a train of $L$ pulses as a block, 
and the working basis is randomly chosen for each block. 
The randomization of overall optical phase is also done for each block of $L$ pulses. 
Bob's receiver is composed of delayed interferometer with its delay being equal to 
the interval $\Delta{\tau}$ of adjacent pulses.  
The longer arm of the interferometer passes through a phase modulator that incurs phase shift 
$\theta_B=$ 0 or $\frac{\pi}{2}$. 
After the interferometer, the pulses are measured by two photon detectors corresponding to bit values ``0'' and ``1''. 
If there is a detection from the superposition of the $l$-th and the $(l-1)$-th original pulses, 
we call it as valid detection at $l$-th timing ($1\leq l\leq L-1$). 

The protocol proceeds as follows, 
which includes predetermined parameters $p_1>0$,  $p_0\coloneqq1-p_1$, $\mu>0$, and $n_{\rm rep}$. 
In its description, $|\bm{\kappa}|$ represents the length of a bit sequence 
$\bm{\kappa}$.
\\
1.~~Alice selects a bit $c\in \{0,1\}$ with probability $p_0$ and $p_1$, 
which correspond to the choice of data basis and check basis, respectively. 
Bob also selects $d\in \{0,1\}$ with probability $p_0$ and $p_1$.\\
2.~~Alice generates $L$ random bits $a_l\in \{0,1\}$ $(0,1,..,L-1)$, and 
prepares 
$L$ optical pulses (system $S$) in the state
 \begin{equation}
\bigotimes_{l=0}^{L-1}\ket{e^{i \theta_l(a_l,c)}\sqrt{\mu}}_{S,l},~~~~
\theta_l(a_l,c)\coloneqq a_l \pi+\frac{\pi}{2}lc, 
\label{prepa}
\end{equation}
where $\ket{\sqrt{\mu}}_{S,l}$ represents coherent state $e^{-\mu/2}\sum_{k}\frac{\mu^{k/2}}{\sqrt{k!}}\ket{k}_{S,l}$ 
of the $l$-th pulse mode. 
Alice randomizes the overall optical phase of the $L$-pulse train, and sends it to Bob. 
\\
3.~~If $d=0$, Bob sets $\theta_B=0$. If $d=1$, he sets $\theta_B=\frac{\pi}{2}$.\\
4.~~If there is no detection of photons in the valid timings, Bob sets $j=0$. 
If the detections have only occurred at a single valid timing, 
the variable $j$ is set to the index of the timing.
If there are detections at multiple timings,
the smallest (earliest) index of them is assigned to $j$. 
If $j\neq 0$, Bob determines his raw key bit $b\in \{0,1\}$ depending on which detector has reported detection at the 
$j$-th timing. 
If both detectors have reported at the $j$-th timing, a random bit is assigned to $b$.
Bob announces $j$ through the public channel.\\ 
5.~~If $j\neq 0$, Alice determines her raw key bit as $a=a_{j-1}\oplus a_j$.\\
6.~~Alice and Bob repeat the above procedures $n_{\rm rep}$ times. 
They publicly disclose $c$ and $d$ for each of the $n_{\rm rep}$ rounds. \\
7-1.~~Alice and Bob define sifted keys 
$\bm{\kappa}_{A1}$ and 
$\bm{\kappa}_{B1}$, respectively, by concatenating their determined bits with $j\neq 0$ and $c=d=1$. They publicly disclose $\bm{\kappa}_{A1}$ and 
$\bm{\kappa}_{B1}$.\\
7-2.~~Alice defines a sifted key 
$\bm{\kappa}_{A0}$ 
by concatenating her determined bits with $j\neq 0$ and $c=d=0$.\\
7-3.~~Bob defines a sifted key 
$\bm{\kappa}_{B0}$ 
by concatenating his determined bits with $j\neq 0$ and $c=d=0$.\\
8.~~Bob corrects the errors in his sifted key $\bm{\kappa}_{B0}$ to make it coincide with Alice's key 
$\bm{\kappa}_{A0}$ through $|\bm{\kappa}_{A0}| f_{\rm EC}$ bits of encrypted public communication from Alice 
by consuming the same length of  
pre-obtained secret key. 
Alice and Bob conduct privacy amplification by shortening their keys 
by $|\bm{\kappa}_{A0}| f_{\rm PA}$ to obtain the final keys. \\

In this paper, we only consider the secure key rate in the asymptotic
limit of an infinite sifted key length. We consider the limit of
$n_{\rm rep}\to \infty$ while the following observed parameters are fixed:
\begin{eqnarray}
Q\coloneqq\frac{|\bm{\kappa}_{A0}|}{n_{\rm rep}p_0^2},~~
E_0\coloneqq\frac{{\rm wt}(\bm{\kappa}_{B0} - 
\bm{\kappa}_{A0})}{n_{\rm rep}p_0^2},
~~E_1\coloneqq\frac{{\rm wt}(\bm{\kappa}_{B1} - 
\bm{\kappa}_{A1})}{n_{\rm rep}p_1^2},
\label{observed}
\end{eqnarray}
where the minus sign is a bit-by-bit modulo-2 subtraction and 
${\rm wt}(\bm{\kappa})$ 
represents the weight,  the number of 1's,
of a bit sequence $\bm{\kappa}$.
In this limit, $f_{\rm EC}$ is given by a function of the bit error rate
$E_0/Q$. In Sec.~\ref{security proof}, the asymptotic value of $f_{\rm PA}$ is determined as a
function of $Q$ and $E_1$. The asymptotic key rate per pulse $R_L$ is then given by
\begin{equation}
R_L= \frac{p_0^2}{L}Q(1-f_{\rm PA}(Q,E_1) -f_{\rm EC} (E_0/Q)).
\label{ki-re-}
\end{equation}
To be precise, the protocol must include the procedure for estimating
the value of $E_0/Q$. We have omitted it in the above protocol because
it is simply done by a random sampling test on the sifted keys 
$\bm{\kappa}_{A0}$
and 
$\bm{\kappa}_{B0}$, and its cost is negligible in the asymptotic key rate.

The security of the above protocol is proved in Sec.~\ref{security proof} under the following 
assumptions on the devices used by Alice and Bob. For clarity, in the main body of 
the paper up to Sec.~\ref{key rates}, we assume that Alice's laser source and modulator produces the states 
in Eq.~(1) precisely. The assumption on the laser will then be relaxed in Sec.~\ref{discussion}.  
The randomization of the overall phase 
in Step 2 is assumed to be done by choosing a common optical phase shift $\phi$ randomly 
from the continuous range of 
 $[0,2\pi)$, and applying it to all the $L$ pulses. 
 This eliminates
 the coherence among different photon-number states. 
The state emitted from Alice in Step 2 is thus expressed as 
\begin{equation}
\sum_{m} \hat{N}_m \big(\bigotimes_{l=0}^{L-1}\ket{e^{i \theta_l(a_l,c)}\sqrt{\mu}}_{S,l}
\bra{e^{i \theta_l(a_l,c)}\sqrt{\mu}}\big)\hat{N}_m, 
\label{enuemu}
\end{equation}
where $\hat{N}_m$ represents the projector onto the subspace with $m$ photons in the $L$ pulses. 

As for Bob's apparatus, we assume that he uses threshold detectors, and further assume that 
 the inefficiency and dark countings of the detectors are 
equivalently represented by an absorber and a stray photon source placed in front of Bob's 
apparatus, and hence they are included in the quantum channel.  
This allows us to regard each of  the detectors in Fig.~\ref{setupfig} as a perfect threshold detector, 
which reports detection if and only if it receives one or more photons. 
To represent a relevant consequence of that assumption in a useful form, we introduce POVM 
(positive operator valued measure) elements for Bob's procedure in Steps 3 and 4. Let 
$\{\hat{B}^{(d)}_j\}_{j=0,...,L-1}$ be the POVM 
for Bob's procedure of determining $j$, when the basis $d$ was selected in Step 1.  
We further decompose the elements for $j\neq 0$ as $\hat{B}^{(d)}_j=\hat{B}^{(d)}_{j,0} + \hat{B}^{(d)}_{j,1}$, where 
$\hat{B}^{(d)}_{j,b}$ corresponds to the outcome $(j,b)$. These operators satisfy
\begin{equation}
\hat{B}^{(d)}_0+\sum_{j=1}^{L-1} 
(\hat{B}^{(d)}_{j,0} + \hat{B}^{(d)}_{j,1}) = \hat{\mathbbm{1}}.
\end{equation}
Under the model of detectors mentioned above, whether there is a detection 
 or not at each timing does not depend on the phase shift applied on the long arm. 
Hence, the procedure to determine $j$ is the same 
for $d=0$ and $d=1$, and we have 
\begin{equation}
\hat{B}_j^{(0)}=\hat{B}_j^{(1)}~~~(0\leq j\leq L-1),
\label{honnnetotatemae}
\end{equation}
which will be used in the security proof given in the next section.

\section{Security proof}\label{security proof}
Here we prove the security of the protocol introduced in Sec.~\ref{protocol} and determine the amount of 
privacy amplification $f_{\rm PA}(Q,E_1)$ in the asymptotic limit. 
Our proof is based on the security analysis with complementarity\cite{2009Koashi} 
as well as the tagging technique
\cite{2004GLLP}. 
For the security proof with complementarity, 
we consider an alternative protocol 
in which Alice's sifted key $\bm{\kappa}_{A0}$ are obtained from measurements on auxiliary qubits 
on a basis ($X$ basis), while Bob, instead of aiming to learn $\bm{\kappa}_{A0}$, 
tries to guess the value of the complementary observable (the outcome of $Y$-basis measurement) for 
Alice's qubits. 
The alternative protocol is designed to fulfill the following conditions:\\\\
  (i) Alice's procedure of releasing optical pulses, making her public announcement 
$\bm{\kappa}_{A1}$, and producing the final key 
  is identical to the actual protocol. 
  \\
  (ii) Bob's procedure of receiving $L$ pulses and making his public announcement $j$ (for each round) 
  and $\bm{\kappa}_{B1}$ 
  in the actual protocol is identical to the corresponding procedure in the alternative protocol. 
  \\\\
 Note that the condition (ii) does not prohibit Bob from making ``additional'' 
 public announcement that is not made in the actual protocol. 
The conditions (i) and (ii) ensure that any attack strategy by Eve in the actual protocol can 
also be applicable to the alternative protocol by ignoring the additional announcement, 
resulting in the identical correlation between Alice's final key and Eve's quantum system.  
Hence, Alice's final key in the actual protocol is secure 
 (random and decoupled from Eve's system) 
if that 
in the alternative protocol is secure against Eve's general attack. 

Now we introduce an alternative protocol satisfying conditions (i) and (ii). 
In the protocol, Alice correlates an auxiliary qubit to each optical pulse. 
We define $X$, $Y$, and $Z$ bases for a qubit as follows. The $Z$ basis is the standard basis 
$\{\ket{0},\ket{1}\}$, 
and a controlled-NOT (CNOT) gate $\hat{U}^{(j)}_{\rm CNOT}$ appearing in the protocol below 
is defined on this basis by 
$\hat{U}^{(j)}_{\rm CNOT} \ket{x}_{A,j}\ket{y}_{A,j-1}
=  \ket{x}_{A,j}\ket{x+y~\rm mod~2}_{A,j-1}$ ($x,y\in \{0,1\}$).
The $X$ basis is $\{\ket{+},\ket{-}\}$, where $\ket{\pm}\coloneqq (\ket{0}\pm \ket{1})/\sqrt{2}$. 
When we represent an outcome of the $X$ basis measurement by a bit, it should be understood that state 
$\ket{+}$ corresponds to bit value 0 and
state $\ket{-}$ to 1. Similarly, we define the $Y$ basis as
$\{\ket{-i},\ket{+i}\}$, with $\ket{\pm i}\coloneqq (\ket{0}\pm i \ket{1})/\sqrt{2}$, 
where we adopt an unconventional rule that  
$\ket{+i}$ corresponds to bit value 1 and 
$\ket{-i}$ to 0 for the convenience of the proof.

The detail of the alternative protocol is described below, where a
step including a different procedure from the actual protocol is marked with an asterisk (*). 
\\

$\bf{Alternative~protocol}.$\\
1.~~Alice selects a bit $c\in \{0,1\}$ with probability $p_0$ and $p_1$, 
which correspond to the choice of data basis and check basis, respectively. 
Bob also selects $d\in \{0,1\}$ with probability $p_0$ and $p_1$.\\
$2^{*}$.~~Alice prepares $L$ auxiliary qubits (system $A$) and $L$ optical pulses (system $S$) in state
\begin{equation}
\ket{\Psi(c)}_{AS}\coloneqq \bigotimes_{l=0}^{L-1}\ket{\psi(c)}_{AS,l}
\label{kprepa}
\end{equation}
depending on her basis choice, where 
\begin{equation}
\ket{\psi(c)}_{AS,l}\coloneqq \frac{1}{\sqrt{2}}(\ket{+}_{A,l}\ket{e^{i\frac{\pi}{2}lc}\sqrt{\mu}}_{S,l}
+\ket{-}_{A,l}\ket{-e^{i\frac{\pi}{2}lc}\sqrt{\mu}}_{S,l}).
\label{replace}
\end{equation} 
She measures the total photon number $m$ in the $L$ pulses with the projective measurement $\{\hat{N}_m\}$, 
and sends the $L$ pulses to Bob. \\
$3^{*}$.~~Bob sets $\theta_B=\frac{\pi}{2}$ regardless of the value of $d$.\\
4.~~If there is no detection of photons in the valid timings, Bob sets $j=0$. 
If the detections have only occurred at a single valid timing, 
the variable $j$ is set to the index of the timing.
If there are detections at multiple timings,
the smallest (earliest) index of them is assigned to $j$. 
If $j\neq 0$, Bob determines his raw key bit $b\in \{0,1\}$ depending on which detector has reported detection at the 
$j$-th timing. 
If both detectors have reported at the $j$-th timing, a random bit is assigned to $b$.
Bob announces $j$ through the public channel. \\
5-$1^{*}$.~~If $j=0$, proceed to Step 6. Otherwise, Alice operates a CNOT  
gate 
$\hat{U}^{(j)}_{\rm CNOT}$ on the $(j-1)$-th qubit (target) and 
the $j$-th qubit (control). \\
5-$2^{*}$.~~Alice measures all the qubits but the $j$-th one on $Z$ basis $\{\ket{0}_{A,l}, \ket{1}_{A,l}\}$ 
to obtain the outcomes $z_l$ ($l\neq j$).
\\
5-$3^{*}$.~~Alice measures the $j$-th qubit on $X$ basis $\{\ket{+}_{A,j}, \ket{-}_{A,j}\}$ 
and determines her raw key bit $a$ accordingly.\\
6.~~Alice and Bob repeat the above procedures $n_{\rm rep}$ times. 
They publicly disclose $c$ and $d$ for each of the $n_{\rm rep}$ rounds. \\
7-1.~~Alice and Bob define sifted keys 
$\bm{\kappa}_{A1}$ and 
$\bm{\kappa}_{B1}$, respectively, by concatenating their determined bits with $j\neq 0$ and $c=d=1$. They publicly disclose $\bm{\kappa}_{A1}$ and 
$\bm{\kappa}_{B1}$.\\
7-2.~~Alice defines a sifted key 
$\bm{\kappa}_{A0}$ 
by concatenating her determined bits with $j\neq 0$ and $c=d=0$.\\
7-$3^*$.~~Bob defines a sifted key 
$\bm{\kappa}^*_{B0}$ 
by concatenating his determined bits with $j\neq 0$ and $c=d=0$.
He publicly discloses $\bm{\kappa}^*_{B0}$.\\
$8^{*}$.~~Alice conducts privacy amplification by shortening her key 
by $|\bm{\kappa}_{A0}| f_{\rm PA}$ to obtain the final key. \\

The above protocol satisfies the condition (ii) because of the following reasons. 
Since Step $3^{*}$ is identical to the actual protocol for $d=1$, so is Bob's announcement of 
$\bm{\kappa}_{B1}$. 
The change in Step $3^{*}$ does not affect the announcement of $j$ in each round due to 
Eq.~(\ref{honnnetotatemae}). 
Note that the change in Step 7-$3^*$ is an additional announcement which is not disclosed 
in the actual protocol. 
In order to see that the condition (i) holds, we will modify the alternative protocol in such a way that 
Alice's procedure dictated in (i) is unchanged. 
Since the outcomes $\{z_l\}$ in Step 5-$2^*$ are neither announced nor used 
in determining the final key, we can omit this step. 
Since a CNOT gate on $Z$ basis is equivalent to a CNOT gate on $X$ basis with target and control exchanged, 
Steps 5-$1^*$ and 5-$3^*$ are equivalently done by measuring all the $L$ qubits on $X$ basis to obtain $L$ bits 
$a_0,a_1,..,a_{L-1}$ as the outcome, and then 
 setting $a=a_{j-1}\oplus a_{j}$.
Since the $X$-basis measurement on all the qubits does not require the knowledge of $j$, 
we may assume that it is done in Step $2^{*}$.  
Then, using the relation 
\begin{equation}
{}_{A,l}\braket{\pm|\psi(c)}_{AS,l}=\frac{1}{\sqrt{2}}\ket{\pm e^{i\frac{\pi}{2}lc}\sqrt{\mu}}, 
\label{puramai}
\end{equation}
we see that the $L$-bit sequence  
 $a_0,a_1,..,a_{L-1}$ is random and 
conditioned on its value the emitted state is identical to Eq.~(\ref{enuemu}). 
Hence, it is equivalent to Steps 2 and 5 of the actual protocol. 
Finally, Steps 7-$3^{*}$ and $8^{*}$ are the same as in the actual protocol as far as Alice is concerned.  
Therefore, the alternative protocol satisfies the condition 
(i), as well as (ii), which means that the security of the alternative protocol implies the 
security of the actual protocol. 

To prove the security of the alternative protocol, we first use the tagging technique 
proposed by Gottesmann $et$ $al.$\cite{2004GLLP}. 
Their idea was to tag the incidents with multiphoton emission, which are totally insecure in the BB84 protocol. 
In a similar vein, we might want to tag the events where the $(j-1)$-th and $j$-th pulses include 
multiphotons upon emission.  
 However, the number of emitted photons in the two pulses is not well-defined due to the phase coherence 
 with other pulses.
Instead, we define a rule to classify tagged $(t=1)$ and untagged $(t=0)$ incidents in terms of 
variables well-defined in the alternative protocol:
\begin{equation}
\sum_{l\neq j} z_l=m~\to~t=0,~~~~\sum_{l\neq j} z_l<m~\to~t=1.
\label{tagrule}
\end{equation}
Let $\bm{\kappa}_{A0,{\rm untag}}$
be the concatenation of all the untagged bits in $\bm{\kappa}_{A0}$, and define the ratio of tagged incidents as 
\begin{equation}
\Delta\coloneqq 1-\frac{ |\bm{\kappa}_{A0,{\rm untag}}|}{ |\bm{\kappa}_{A0}|}.
\label{Delta}
\end{equation}
According to \cite{2004GLLP}, if a sufficient amount of privacy amplification on
$\bm{\kappa}_{A0,{\rm untag}}$ to make it secure is given by $|\bm{\kappa}_{A0,{\rm untag}}|g_{\rm PA}(Q,E_1,\Delta)$, 
$\bm{\kappa}_{A0}$ can be made to be secure by reducing its length by 
$|\bm{\kappa}_{A0}|f_{PA}$ if it satisfies
\begin{equation}
f_{\rm PA}(Q,E_1)\geq \mathop{\rm max}_{\Delta}\big(\Delta+(1-\Delta)g_{\rm PA}(Q,E_1,\Delta)\big).
\label{fPA}
\end{equation}

Let us discuss the implication of the condition Eq.~(\ref{tagrule}) for the tagging, 
and derive important relations that will be used in the subsequent proof of
security.
According to Eq.~(\ref{replace}), it is not difficult to see that 
${}_{A,l}\braket{0|\psi(c)}_{AS,l}$ includes only even number of photons, and 
${}_{A,l}\braket{1|\psi(c)}_{AS,l}$ does odd number of photons. 
For convenience, let us define projectors related to such a property by
\begin{equation}
\hat{\Upsilon}_{AS}:=\bigotimes_{l=0}^{L-1} \hat{\Upsilon}^{(l)},
\;\;
 \hat{\Upsilon}^{(l)}:= \hat{P}(\ket{0}_{A,l})
\left(
\sum_{n: {\rm even}} \hat{P}(\ket{n}_{S,l})
\right)
+
\hat{P}(\ket{1}_{A,l})
\left(
\sum_{n: {\rm odd}} \hat{P}(\ket{n}_{S,l})
\right),
\\
\end{equation}
where $\hat{P}(\ket{\cdot})=\ket{\cdot}\bra{\cdot}$. Notice that the initial state in Eq.~(\ref{kprepa}) satisfies
\begin{equation}
 \hat{\Upsilon}_{AS} \ket{\Psi(c)}_{AS}=\ket{\Psi(c)}_{AS}.
 \label{korekore}
\end{equation}
Thanks to the correlation specified by $\hat{\Upsilon}_{AS}$, the measured
quantities $\{z_l\}$ are related to the parity of the photon numbers
in the system $S$. To see this, let us define the projector corresponding to 
the state of $m_l$ photons in the $l$-th pulse by  
\begin{equation}
 \hat{N}_{\{m_l\}}\coloneqq \bigotimes_{l=0}^{L-1} \hat{P}(\ket{m_l}_{S,l}).
\end{equation}
Alice's procedure of determining $\{z_l\}~(l\neq j)$ at Steps 
5-1$^*$ and 5-2$^*$ will be associated with the projector defined by
\begin{eqnarray}
 \hat{F}^{(j)}_{\{z_l\}}&\coloneqq&
\hat{U}_{\rm CNOT}^{(j)\dagger} \left(\hat{\mathbbm{1}}_{A,j}\otimes \Big(
\bigotimes_{l\neq j} \hat{P}(\ket{z_l}_{A,l})\Big)\right)
\hat{U}_{\rm CNOT}^{(j)} \nonumber \\
&=&\left[\hat{P}(\ket{0}_{A,j-1}\ket{z_{j-1}}_{A,j})
+\hat{P}(\ket{1}_{A,j-1}\ket{1-z_{j-1}}_{A,j})\right]
\bigotimes_{l\neq j-1,j} \hat{P}(\ket{z_l}_{A,l}).
\end{eqnarray}
Then, it is easy to confirm that 
\begin{eqnarray}
 &&(\hat{F}^{(j)}_{\{z_l\}}\otimes \hat{N}_{\{m_l\}})\hat{\Upsilon}_{AS} \neq 0 \nonumber \\
&&{\rm only~if}~~
z_l = m_l~~{\rm mod}~2~(l\neq j-1,j)
~~{\rm and}~~
z_{j-1}= m_{j-1}+m_j~~{\rm mod}~2.
\end{eqnarray}
Since $\hat{N}_m\hat{N}_{\{m_l\}}=0$ unless $\sum_{l} m_l =  m$, we have
\begin{eqnarray}
&& (\hat{F}^{(j)}_{\{z_l\}}\otimes \hat{N}_m\hat{N}_{\{m_l\}})\hat{\Upsilon}_{AS} \neq 0 \nonumber \\
&&{\rm only~if}~~
z_l \le  m_l~(l\neq j-1,j),~~
z_{j-1}\le  m_{j-1}+m_j 
~~{\rm and}~~
\sum_{l} m_l =  m.
\end{eqnarray}
If we confine ourselves to the case with $\sum_{l\neq j} z_l =  m$, the
condition in the above equation is satisfied only by
$z_l =  m_l~(l\neq j-1,j)$ and $z_{j-1}=  m_{j-1}+m_j$.
We thus conclude that
\begin{equation}
 (\hat{F}^{(j)}_{\{z_l\}}\otimes \hat{N}_m)\hat{\Upsilon}_{AS} 
= (\hat{F}^{(j)}_{\{z_l\}}\otimes
 \hat{\Xi}^{(j)}_{\{z_l\}})\hat{\Upsilon}_{AS}
~~{\rm for}~~ 
\sum_{l\neq j} z_l =  m,
\label{operaza}
\end{equation}
where
\begin{eqnarray}
 \hat{\Xi}^{(j)}_{\{z_l\}} &:=& 
\hat{P}(\ket{0}_{S,j-1}\ket{0}_{S,j})
\bigotimes_{l\neq j-1,j} \hat{P}(\ket{z_l}_{S,l})
~~{\rm for}~~
z_{j-1}=0 
\label{Xi0}
\\
\hat{\Xi}^{(j)}_{\{z_l\}} &:=& 
[\hat{P}(\ket{0}_{S,j-1}\ket{1}_{S,j})
+\hat{P}(\ket{1}_{S,j-1}\ket{0}_{S,j})
]
\bigotimes_{l\neq j-1,j} \hat{P}(\ket{z_l}_{S,l})
~~{\rm for}~~
z_{j-1}=1.
\label{Xi1}
\end{eqnarray}
This may lead to an interpretation that, whenever the event is untagged, 
every pulse should have 
contained no more than one photon upon emission, and the $(j-1)$-th and
the $j$-th pulse pair contained no more than one photon in total. On the
other hand, we should also take notice that 
Alice's measurement of $\{z_l\}$ ($l\neq j$) in the alternative protocol
 can be carried out only after the pulse train was measured by Bob and
 the value of $j$ was announced. Hence the above interpretation has an
 ambiguity in the operational sense, which is why we only use strict
 mathematical statements of Eqs.~(\ref{korekore}) and (\ref{operaza}) in the subsequent proof
 and do not rely on the interpretation.

Our next goal is to 
determine the amount of privacy amplification $g_{PA}(Q,E_1,\Delta)$ to make the untagged portion of the key secure, 
 using the proof based on complementarity\cite{2009Koashi}. 
In that proof, we consider a measurement which is complementary to 
the measurement on the qubits to determine the sifted key 
$\bm{\kappa}_{A0,\rm untag}$. Let us introduce the following procedure instead of the 
Steps 5-$3^*$ and 7-2.
\\\\
5-$3^{**}$. If $c=1$, 
Alice measures the $j$-th qubit on $X$ basis $\{\ket{+}_{A,j}, \ket{-}_{A,j}\}$ 
and determines her raw key bit $a$ accordingly.
If $c=0$, 
Alice measures the $j$-th qubit on $Y$ basis $\{\ket{-i}_{A,j}, \ket{+i}_{A,j}\}$ 
and determines her raw key bit $a$ accordingly.\\
7-$2^{**}$. Alice defines a sifted key 
$\bm{\kappa}^*_{A0}$ 
by concatenating her determined bits with $j\neq 0$ and $c=d=0$.
\\\\
Suppose that we have a bound 
$\delta_{{\rm untag}}(Q,E_1,\Delta)$, 
which asymptotically satisfies 
\begin{equation}
\delta_{\rm untag}(Q,E_1,\Delta) \ge \frac{{\rm wt} 
(\bm{\kappa}^*_{B0,{\rm untag}}
-\bm{\kappa}^*_{A0,{\rm untag}})}{|\bm{\kappa}^*_{A0,{\rm untag}}|},
\end{equation}
where $\bm{\kappa}^*_{A0,{\rm untag}}$ and $\bm{\kappa}^*_{B0,{\rm untag}}$ are 
the concatenations of all the untagged bits in $\bm{\kappa}^*_{A0}$ and $\bm{\kappa}^*_{B0}$, respectively.
Notice that the measurement on Alice's qubits for extracting $\bm{\kappa}_{A0}$ or $\bm{\kappa}^*_{A0}$ can 
be postponed until after Step 7-$3^*$, namely, after she learns the values of 
$Q, E_1, \Delta$  and $\bm{\kappa}^*_{B0,\rm untag}$. 
Then, an extreme case of 
$\delta_{{\rm untag}}(Q,E_1,\Delta)=0$ will mean that the state of 
 $|\bm{\kappa}_{A0,{\rm untag}}|$
untagged qubits before the measurement is exactly a $Y$-basis eigenstate specified by 
$\bm{\kappa}^*_{B0,{\rm untag}}$, 
and hence $\bm{\kappa}_{A0,{\rm untag}}$, which is an outcome of $X$-basis measurement, 
is secure (random and decoupled from Eve's system). 
For a nonzero value of $\delta_{{\rm untag}}(Q,E_1,\Delta)$, 
we need a privacy amplification to make it secure, 
and a sufficient amount in the asymptotic limit was derived in \cite{2009Koashi} as
\begin{equation}
g_{\rm PA}(Q,E_1,\Delta)=h(\delta_{\rm untag}(Q,E_1,\Delta))
\label{kirekire}
\end{equation}
for $\delta_{\rm untag}(Q,E_1,\Delta)\leq 1/2$, 
where $h(x)\coloneqq -{\rm log}_2 x-(1-x){\rm log}_2(1-x)$ represents the binary entropy function.

It can be shown that 
 $\delta_{\rm untag}$ is connected
to the check-basis error rate $E_1$ of the actual protocol through random sampling. 
For given values of $c$ and $j$, 
Alice's procedure of determining 
$\{z_l\}$ and $a$ at Steps 5-$1^*$, 5-$2^*$
and 5-$3^{**}$ corresponds to the projection onto the state $\ket{\mathcal{A}_{a,\{z_l\}}^{(c,j)}}_A$, 
which is defined by 
\begin{equation}
\ket{\mathcal{A}_{a,\{z_l\}}^{(c,j)}}_A
\coloneqq
\frac{1}{\sqrt{2}}
 \hat{U}^{(j)\dagger}_{\rm CNOT} \left(
\Big(\ket{0}_{A,j}-(-1)^a~i^{~c+1} \ket{1}_{A,j} 
\Big)
\bigotimes_{l\neq j}\ket{z_l}_{A,l}\right).\nonumber
\end{equation}
Since these states satisfy 
\begin{equation}
\hat{F}_{\{z_l\}}^{(j)}\ket{\mathcal{A}_{a,\{z_l\}}^{(c,j)}}_A=\ket{\mathcal{A}_{a,\{z_l\}}^{(c,j)}}_A,
\label{kaijin}
\end{equation}
Eqs.~(\ref{operaza}) and (\ref{kaijin}) 
lead to 
\begin{equation}
{}_A\bra{\mathcal{A}_{a,\{z_l\}}^{(c,j)}} \hat{N}_m \hat{\Upsilon}_{AS}=
{}_A\bra{\mathcal{A}_{a,\{z_l\}}^{(c,j)}} \hat{\Xi}_{\{z_l\}}^{(j)} \hat{\Upsilon}_{AS}
~~{\rm for}~~\sum_{l\neq j}z_l=m. 
\label{wakazono}
\end{equation}
From Eq.~(\ref{korekore}), we have 
\begin{equation}
{}_A\bra{\mathcal{A}_{a,\{z_l\}}^{(c,j)}} \hat{N}_m \ket{\Psi(c)}_{AS}=
{}_A\bra{\mathcal{A}_{a,\{z_l\}}^{(c,j)}} \hat{\Xi}_{\{z_l\}}^{(j)} \ket{\Psi(c)}_{AS}
~~{\rm for}~~\sum_{l\neq j}z_l=m.
\label{ningyou}
\end{equation}
 The basis-choice dependence of states $\ket{\mathcal{A}_{a,\{z_l\}}^{(c,j)}}_A$ and  
$\ket{\Psi(c)}_{AS}$ can be represented by
\begin{equation}
 \ket{\mathcal{A}_{a,\{z_l\}}^{(c,j)}}_A=\Big(\hat{P}(\ket{0}_{A,j})+ i^{~c}~\hat{P}(\ket{1}_{A,j})\Big)
 \ket{\mathcal{A}_{a,\{z_l\}}^{(0,j)}}_A
\label{Anituite}
\end{equation}
and
\begin{equation}
 \ket{\Psi(c)}_{AS}=\left(\bigotimes_{l=0}^{L-1} i^{~l\hat{m_l}c}
\right)\ket{\Psi(0)}_{AS},
\label{Psinituite}
\end{equation}
where $\hat{m}_l:=\sum_m m \hat{P}(\ket{m}_l)$ is the photon number operator
for the $l$-th pulse.
Since the range of the projector $\hat{\Xi}_{\{z_l\}}^{(j)}$ includes only zero- or one-photon
states for each mode, we have
\begin{equation}
 [(\hat{P}(\ket{0}_{A,j})+ (-i)^c \hat{P}(\ket{1}_{A,j}))\otimes \hat{\Xi}_{\{z_l\}}^{(j)}] \hat{\Upsilon}_{AS}
= (-i)^{c \hat{m}_j} \hat{\Xi}_{\{z_l\}}^{(j)} \hat{\Upsilon}_{AS}.
\label{Upsilonnituite}
\end{equation}
Combining  Eqs.~(\ref{korekore}), (\ref{Anituite}), (\ref{Psinituite}) and 
(\ref{Upsilonnituite}), we obtain
\begin{equation}
 {}_A\bra{\mathcal{A}_{a,\{z_l\}}^{(c,j)}}\hat{\Xi}_{\{z_l\}}^{(j)}\ket{\Psi(c)}_{AS}
={}_A\bra{\mathcal{A}_{a,\{z_l\}}^{(0,j)}} (-i)^{\hat{m}_jc}\left(\bigotimes_{l=0}^{L-1} i^{~l\hat{m_l}c}
\right)   \hat{\Xi}_{\{z_l\}}^{(j)} \ket{\Psi(0)}_{AS}.
\label{30}
\end{equation}
Using the definition of Eqs.~(\ref{Xi0}) and (\ref{Xi1}), it is easy to confirm that
\begin{equation}
 (-i)^{c \hat{m}_j}\left(\bigotimes_{l=0}^{L-1} i^{~l\hat{m_l}c}
\right)   \hat{\Xi}_{\{z_l\}}^{(j)}
= i^{~(j-1)z_{j-1} c} \left(\prod_{l\neq j-1,j} i^{~lz_l c}\right)
\hat{\Xi}_{\{z_l\}}^{(j)}
\label{31}
\end{equation}
holds. 
Therefore,  we have
\begin{equation} 
{}_A\bra{\mathcal{A}_{a,\{z_l\}}^{(0,j)}}
\hat{\Xi}_{\{z_{j-1}\}}^{(j)}\ket{\Psi(0)}_{AS}
=(-i)^{u(j)}{}_A\bra{\mathcal{A}_{a,\{z_l\}}^{(1,j)}}
\hat{\Xi}_{\{z_{j-1}\}}^{(j)}\ket{\Psi(1)}_{AS},
\label{awaseru}
\end{equation} 
where $u(j)\coloneqq \sum_{l\neq j-1,j}l z_l+(j-1)z_{j-1}$ and this leads, with Eq.~(\ref{ningyou}), to  
\begin{equation} 
{}_A\bra{\mathcal{A}_{a,\{z_l\}}^{(0,j)}}
\hat{N}_m\ket{\Psi(0)}_{AS}
=(-i)^{u(j)}{}_A\bra{\mathcal{A}_{a,\{z_l\}}^{(1,j)}}
\hat{N}_m\ket{\Psi(1)}_{AS}
~~{\rm for}~~\sum_{l\neq j}z_l=m.
\label{hitoshi}
\end{equation}
This relation may suggest that for untagged incidents, the state of pulses transmitted from Alice would be 
independent of the value of $c$, and hence the $c=d=1$ incidents would be regarded as a fair sampling. 
Again, this interpretation suffers from ambiguity 
since the protocol assumes that Alice's qubits are 
measured only after the optical pulses are received by Bob and 
the value of $j$ is announced. 
Therefore we need a mathematical proof for the fairness of the sampling, 
which is given in Appendix A. The proof confirms that
\begin{equation}
\frac{{\rm wt} 
(\bm{\kappa}^*_{B0,{\rm untag}}
-\bm{\kappa}^*_{A0,{\rm untag}})}{{\rm wt} 
(\bm{\kappa}_{B1,{\rm untag}}
-\bm{\kappa}_{A1,{\rm untag}})}
=\left(\frac{p_0}{p_1}\right)^2
\label{ransam}
\end{equation}
holds in the limit of $n_{\rm rep}\to \infty$. 
Then we have 
\begin{eqnarray}
\frac{{\rm wt} 
(\bm{\kappa}^*_{B0,{\rm untag}}
-\bm{\kappa}^*_{A0,{\rm untag}})}{|\bm{\kappa}^*_{A0,{\rm untag}}|}
&=&
\left(\frac{p_0}{p_1}\right)^2\frac{{\rm wt} 
(\bm{\kappa}_{B1,{\rm untag}}
-\bm{\kappa}_{A1,{\rm untag}})}
{|\bm{\kappa}^*_{A0,{\rm untag}}|}\nonumber \\ 
&\leq& \left(\frac{p_0}{p_1}\right)^2
\frac{{\rm wt} 
(\bm{\kappa}_{B1}
-\bm{\kappa}_{A1})}
{|\bm{\kappa}^*_{A0,{\rm untag}}|}
=\frac{E_1}{Q(1-\Delta)}.
\end{eqnarray}
We thus conclude that asymptotically a privacy amplification with a ratio
\begin{equation}
g_{PA}(Q,E_1,\Delta)=h\Big(\frac{E_1}{Q(1-\Delta)}\Big)
\label{modokashii}
\end{equation}
is enough to make the untagged portion of the sifted key secure. 

Since the argument of the max in Eq.~(\ref{fPA}) with Eq.~(\ref{modokashii}) is an increasing
function of $\Delta$ for $0\le \Delta \le 1-(2E_1/Q)$, $f_{\rm PA}$ will
be determined through finding an upperbound on $\Delta$. 
According to the definition of Eq.~(\ref{Delta}), what we need is a lower bound on
$|\bm{\kappa}_{A0,{\rm untag}}|$, which is determined as follows. 
If we denote by $n(condition)$ the
number of rounds satisfying the {\it condition} in the $n_{\rm rep}$ rounds
repeated in the alternative protocol, we have 
$|\bm{\kappa}_{A0}|=n(c=d=0,j\neq 0)$ and 
$|\bm{\kappa}_{A0,{\rm untag}}|=n(c=d=0,j\neq 0, t=0)$,
where $t=0$ is equivalent to  $\sum_{l\neq j} z_l=m$ 
according to Eq.~(\ref{tagrule}).
Under a given attack strategy of Eve,
the statistics of $n(c=d=0,j\neq 0)$ and
$n(c=d=0,j\neq 0, t=0)$ is unchanged if 
we omit Step 5-3$^*$ and stop the protocol at Step 6. 
We may further equivalently replace Steps 5-1$^*$ and 5-2$^*$ with 
a procedure of
measuring the $L$ qubits on the $Z$ basis $\{\ket{0}_{A,l},\ket{1}_{A,l}\}$ to obtain the
outcomes $z'_0,\cdots z_{L-1}'$, followed by substitutions
$z_l:=z_l'$ $(l\neq j-1,j)$ and $z_{j-1}:=z_{j-1}'+z_j'~~{\rm mod}~~2$ in case
of $j\neq 0$. 
Let us define a set of values of $L$ nonnegative integers as 
\begin{equation}
 \Gamma^{(m)}:=\{(k_0,\cdots, k_{L-1})| k_{l-1}+k_l \le
  1 (1\le l \le L-1), \sum_{l=0}^{L-1} k_l=m \},
\end{equation}
and operators associated with it by
\begin{equation}
 \hat{\Pi}^{(m)}_{A}:= \sum_{\{z'_l\}\in \Gamma^{(m)}} \bigotimes_{l=0}^{L-1}
  \hat{P}(\ket{z'_l}_{A,l}), \;\;
\hat{\Pi}^{(m)}_{S}:= \sum_{\{m_l\}\in \Gamma^{(m)}} \bigotimes_{l=0}^{L-1}
  \hat{P}(\ket{m_l}_{S,l}).
\end{equation}
We see that $(z'_0,\cdots z_{L-1}')\in \Gamma^{(m)}$ implies $\sum_{l\neq j} z_l=m$
regardless of the value of $j$, as long as $j\neq 0$. Hence we have 
\begin{eqnarray}
 &&n(c=d=0,j\neq 0, t=0)\nonumber \\
 &\ge& n(c=d=0,j\neq 0, (z'_0,\cdots z_{L-1}')\in \Gamma^{(m)})
\nonumber \\
&=& n(c=d=0,j\neq 0)-n(c=d=0,j\neq 0, (z'_0,\cdots z_{L-1}')\notin \Gamma^{(m)})
\nonumber \\
&\ge& n(c=d=0,j\neq 0) - n(c=d=0, (z'_0,\cdots z_{L-1}')\notin \Gamma^{(m)}).
\label{4array}
\end{eqnarray}
The number $n(c=d=0, (z'_0,\cdots z_{L-1}')\notin \Gamma^{(m)})$ is independent of Eve's
strategy, and it follows the binomial distribution with success probability 
 $p_0^2 r_{\rm tag}$ with
\begin{equation}
 r_{\rm tag}\coloneqq 1-\sum_m {}_{AS}\bra{\Psi(0)}\hat{\Pi}^{(m)}_{A}\otimes \hat{N}_m
\ket{\Psi(0)}_{AS}.
\label{tokkyu}
\end{equation}
Since $z'_l=m_l~~{\rm mod}~~2$ and $(m_0,\ldots,m_{L-1})\in \Gamma^{(m)}$
imply $(z'_0,\cdots z_{L-1}')\in \Gamma^{(m)}$, we have 
$\hat{\Pi}^{(m)}_{S}\hat{\Upsilon}_{AS}=
(\hat{\Pi}^{(m)}_{A}\otimes\hat{\Pi}^{(m)}_{S})\hat{\Upsilon}_{AS}$.
On the other hand, $z'_l=m_l~~{\rm mod}~2$ and 
$\sum_l z'_l=\sum_l m_l$ imply $z'_l=m_l$, which leads to 
$(\hat{\Pi}^{(m)}_{A}\otimes \hat{N}_m)\hat{\Upsilon}_{AS}
=(\hat{\Pi}^{(m)}_{A}\otimes \hat{\Pi}^{(m)}_{S})\hat{\Upsilon}_{AS}$.
We thus obtain
\begin{equation}
 (\hat{\Pi}^{(m)}_{A}\otimes \hat{N}_m)\hat{\Upsilon}_{AS}
 =\hat{\Pi}^{(m)}_{S}\hat{\Upsilon}_{AS}.
 \label{regardless}
\end{equation}
Combined with Eq.~(\ref{korekore}), we obtain
\begin{equation}
 r_{\rm tag}=1-\sum_m {}_{AS}\bra{\Psi(0)}
 \hat{\Pi}^{(m)}_{S}
\ket{\Psi(0)}_{AS},
\label{tagtag}
\end{equation}
which gives us a clear interpretation of quantity 
$r_{\rm tag}$ being the probability that the $L$-pulse train emitted
from Alice contains at least two photons in the same pulse or in 
neighboring pulses. As a function of $\mu$, it is calculated as
\begin{eqnarray}
 r_{\rm tag} &=& 1- \sum_{m=0}^{\lceil L/2 \rceil}
e^{-\mu L}\mu^m
\frac{(L+1-m)!}{m!(L+1-2m)!}.
\label{rtagmu}
\end{eqnarray}
In the asymptotic limit of $n_{\rm rep}\to \infty$, Eq.~(\ref{4array}) implies 
\begin{equation}
 \frac{n(c=d=0,j\neq 0, t=0)}{n_{\rm rep}}
\ge \frac{n(c=d=0,j\neq 0)}{n_{\rm rep}} - p_0^2 r_{\rm tag},
\end{equation}
which means that 
$|\bm{\kappa}_{A0,{\rm untag}}|/n_{\rm rep}
\ge |\bm{\kappa}_{A0}|/n_{\rm rep}
- p_0^2 r_{\rm tag}$. Using  Eqs.~(\ref{observed}) and (\ref{Delta}), we have
\begin{equation}
 \Delta \le \frac{r_{\rm tag}}{Q}.
\end{equation}
Hence, for $r_{\rm tag}\le Q-2E_1$, choosing 
\begin{equation}
f_{\rm PA}(Q,E_1)= \frac{r_{\rm tag}}{Q} 
+\left(1- \frac{r_{\rm tag}}{Q}\right)
h\left(\frac{E_1}{Q-r_{\rm tag}}\right)
\end{equation}
makes the alternative protocol, and hence the actual protocol, secure.
An achievable asymptotic key rate per pulse is thus given by
\begin{equation}
R_L=
\frac{p_0^2}{L}\Big(( Q-r_{\rm tag}) (1-h\Big( \frac{E_1}{Q-r_{\rm tag}}\Big))
-Qf_{EC}(E_0/Q)\Big)
\label{ineq}
\end{equation}
whenever the right-hand side is positive.

 \section{Key rates}\label{key rates}
\begin{figure}[htbp]
\begin{center}
  \raisebox{10mm}{\includegraphics[width=100mm]{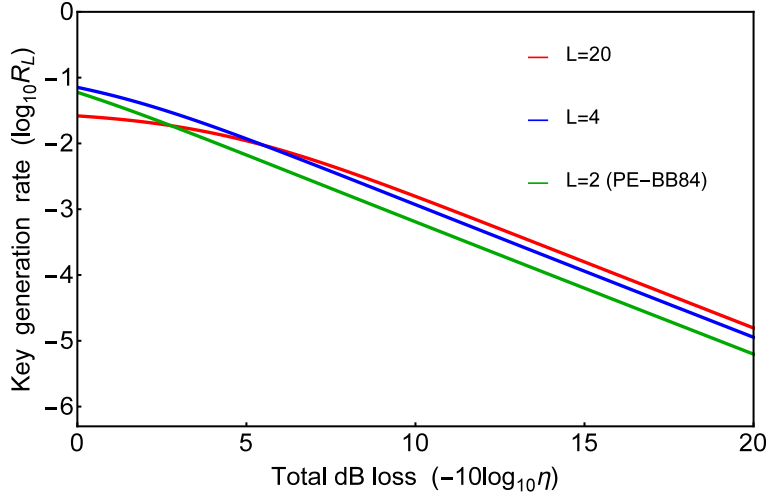}}
  \caption{Secure key rate per pulse $R_L$ as a function of the overall channel transmission $\eta$. 
   The bit error rate of the sifted key 
  is fixed to be 3\% for both bases. 
  The block size $L$ is chosen to be 2, 4, and 20, 
  where $L=2$ corresponds to the PE-BB84 protocol and the other values to the DQPS protocol.
}
  \label{QIPyou}
 \end{center}
\end{figure}

 We show results of numerical calculation of the key rate per pulse $R_L$ 
 given by Eq.~(\ref{ineq}) to compare the conventional passive PE-BB84 protocol ($L$=2)  
 and the DQPS protocol ($L\geq 3$). In Fig.~\ref{QIPyou}, dependence of $R_L$ on 
 overall transmission $\eta$ 
 (including detector efficiency) is shown for $L=2,4,20$. 
 For channel with transmission $\eta$, we assumed 
 $Q= 1-e^{-(L-1) \mu \eta}$, reflecting the fact that there are $(L-1)$ valid timings per block of pulses. 
  The observed error rates of the sifted key $E_0/Q$ and $E_1/Q$ were both set to be $3\%$, 
  regardless of $\eta$ and $\mu$. 
  We also adopted $f_{\rm EC}(E_0/Q)=h(E_0/Q)$ and  
  $p_0= 1$. 
  The key rate $R_L$ was then optimized over $\mu$ for each value of $\eta$.
  From Fig.~\ref{QIPyou}, we see that, $R_L$ for different values of $L$ are all proportional to $\eta^2$ 
 in the limit of small $\eta$, but its coefficient increases as $L$ gets larger. 
 For example, at $-10{\rm log}_{10}\eta=20$, we found that 
 $R_{20}/R_{2}\cong 2.67$, 
 which clearly shows an advantage of the DQPS protocol 
 over the PE-BB84 protocol when we use essentially the same hardware. 
We also see that even in the limit of no loss $(\eta \to 1)$, the DQPS protocol with $L=4$ is superior to 
the PE-BB84 protocol. 

\section{Discussion}\label{discussion}
Figure 2 shows that the optimized key rates are proportional to $\eta^2$ in the limit of $\eta \to 0$, 
with its coefficient dependent on the block size $L$. 
In the special case where the bit error rate is zero, we can analytically determine the coefficient 
as a function of $L$. 
For $L\mu^2 \ll 1$, the parameter $r_{\rm tag}$ in Eq.~(\ref{rtagmu}) is approximated as 
$r_{\rm tag}=\frac{3L-2}{2}\mu^2$. For $L\mu \eta \ll 1$, the parameter $Q$ is approximated as $Q=(L-1)\mu \eta$. 
Hence, for $L\eta^2 \ll 1$, the key rate $R_L=(Q-r_{\rm tag})/L$ is optimized at 
$\mu=\mu^{\rm opt}\coloneqq \frac{L-1}{3L-2}\eta$ to attain the optimal value 
$R_L^{\rm opt}\coloneqq \frac{(L-1)^2}{2L(3L-2)}\eta^2$. 
In the limit of a large block size, we have 
$R_{L\to \infty}^{\rm opt}=\eta^2/6$ and $R_{L\to \infty}^{\rm opt}/R_2^{\rm opt}=8/3$. 
The result seems interesting in the sense that the secure key rate 
for a large value of $L$ 
 is $more$ $than$ $twice$
as large as that of $L=2$ while the inherent loss in the passive interferometer 
itself is 1/2 for $L=2$. 
On the other hand, it does not mean that the key rate exceeds the case of $L=2$ without 
the interferometer loss, namely, implementation with an ideal active optical switch. 
Since $R_L^{\rm opt} \propto \eta^2$ holds in the limit of small $\eta$, 
the key rate of an ideal active protocol is 4 times the rate of the passive one for $L=2$. 
If the loss in the optical switch is taken into account, 
the passive DQPS protocol is more efficient than the active PE-BB84 protocol 
when the loss of optical switch is larger than $\sim 20\%$. 

While we have assumed so far that the initial pure state represented in 
Eq.~(\ref{prepa}) is prepared by Alice, the proof can be 
extended to a general light source, which is shown in Appendix B. The proof there assumes that 
the phase modulator (PM in Fig.~\ref{setupfig}) works perfectly, and that 
every $L$-pulse train from the source is independent and represented by the same density operator
 $\hat{\sigma}_S$ (not necessarily identical for each pulse). 
For the general light source described above, 
the secure key rate is still given by Eq.~(\ref{ineq}) 
with 
\begin{equation}
r_{\rm tag}=1-\sum_m {\rm tr}\left( \hat{\Pi}^{(m)}_{S}
\hat{\sigma}_{S}\right). 
\end{equation}

Even when the state $\hat{\sigma}_S$ of the $L$ pulse train is unknown, 
an upper bound on $r_{\rm tag}$ can be determined from an off-line coincidence measurement 
on the light source using a few detectors. 
As shown in Appendix C, the calibration method reveals an upper bound 
that is close to the true value of $r_{\rm tag}$, as long as the state from the source is close 
to a coherent state with its mean photon number $\mu \ll L^{-1/2}$.

Although the key rate of the DQPS protocol can be improved by using decoy-state method, 
it is less effective as $L$ gets larger, because only the statistics of the total number 
of photons emitted in the $L$ pulses  are obtained and no further information on their distribution 
over the $L$ pulses is available. 
On the other hand, the decoy state BB84 protocol uses the knowledge on the probability 
of higher photon numbers from the 
light source, 
which will require more complicated devices for calibration. 
Since the calibration of the light source in the DQPS protocol is almost as simple as that of PE-BB84, 
the DQPS protocol 
will be useful for the practical cases where one prefers a simple setup for short 
distance communication\cite{2013Richard}.


Another possible improvement of our result may be obtained from the expected 
 robustness of general DPS protocols against PNS attacks. 
In the DPS protocols (including the DQPS protocol), Eve's attempts to control the timing of detection $j$ 
tends to increase the probability of a bit error, 
which is expected to result in the robustness against PNS attacks. 
In a security proof of the DPS protocol \cite{Tamaki2012}, the robustness can be seen as a 
$\eta^{\frac{3}{2}}$-dependence of the key rate 
in the range of small $\eta$. 
In contrast, our key rate of the DQPS protocol scales as $\eta^2$.
This is because 
our proof assumed the pessimistic assumption that 
Eve is able to control the value of $j$ without causing any bit error. 
If we analyze the security based on the proof technique for the DPS protocol\cite{Tamaki2012},
 our protocol may benefit from the robustness against PNS attacks without using decoy states. 
 
As a conclusion, 
we have proved the security of differential quadrature phase shift (DQPS) quantum key distribution protocol, 
which can be implemented with almost the same setup as the phase-encoding (PE) BB84 protocol. 
The proof is based on the a careful adaptation of the tagging idea and the complementarity argument. 
We found that the key generation rate exceeds that of the PE-BB84 protocol for any channel transmission, 
and is 8/3 as high as the rate of the PE-BB84 protocol in the limit of small transmission. 

\section*{Acknowledgement}
We thank Hiroki Takesue, Yasunari Suzuki, Zhiyuan Tang and Hoi-Kwong Lo for helpful discussions. 
This work was
supported by the ImPACT Program of the Council for Science, Technology and
Innovation (Cabinet Office, Government of Japan) and the Photon Frontier Network
Program (MEXT).

\appendix
\section{Untagged check-basis outcomes as an unbiased sample}
Here, we prove Eq.~(\ref{ransam}) in the main text by showing 
that the untagged rounds with $c=1$ can be regarded as a random sample extracted from the whole untagged events.
For fixed $c,j~(\neq 0)$ and $m$, 
define a projector 
$\hat{T}^{(c,j,m)}_{a,t}
\coloneqq 
\sum_{\{z_l\}}\ket{\mathcal{A}_{a,\{z_l\}}^{(c,j)}}_A\bra{\mathcal{A}_{a,\{z_l\}}^{(c,j)}}$
~~where the summation is over $\{z_l\}$ satisfying 
$\sum_{l\neq j} z_l=m$ for $t=0$ and 
$\sum_{l\neq j} z_l<m$ for $t=1$. 
The projector $\hat{T}^{(c,j,m)}_{a,t}$ can be regarded as the POVM element for the measurement on system $A$ 
to determine $a$ and $t$ through Steps  5-$1^*$, 5-$2^*$, and 5-$3^{**}$ with the rule of Eq.~(\ref{tagrule}). 
Although the protocol does not define the values of $a$, $b$, and $t$ in case of $j=0$, 
it simplifies the notations if we also define those values to be 
$a=b=t=0$ for $j=0$, and define 
$\hat{T}^{(c,0,m)}_{a,t}$ accordingly.
We label each of the $n_{\rm rep}$ rounds by $r=1,2,\ldots ,n_{\rm rep}$, 
and use $c_r,a_r,b_r,j_r,m_r,t_r$ to denote the values of $c,a,b,j,m,t$ in the $r$-th round. Let 
$\bm{c},\bm{a},\bm{b},\bm{j},\bm{m},\bm{t}$ be vectors with $n_{\rm rep}$ elements corresponding to 
$r=1,2,\ldots ,n_{\rm rep}$.
With these notations, the procedure of determining these vectors in the alternative protocol 
(with replacement 5-$3^{**}$) is summarized as follows. \\
i)~~Alice selects $\bm{c}$ randomly, prepares 
$\hat{\rho}_{AS}(\bm{c})\coloneqq \bigotimes_{r=1}^{n_{\rm rep}} \hat{\sigma}_{AS}(c_r)$
with $\hat{\sigma}_{AS}(c_r) \coloneqq \ket{\Psi(c_r)}_{AS}\bra{\Psi(c_r)}$, 
and measures $\bm{m}$ by a projection measurement.\\
ii)~~Eve's attack on $n_{\rm rep}$ copies of system $S$ followed by Bob's measurement determines $\bm{j}$ and 
$\bm{b}$. 
For a given attack strategy by Eve, this whole procedure on $n_{\rm rep}$ systems should be represented by 
POVM with elements $\{\hat{D}_{\bm{j},\bm{b}}\}$. \\
iii)~~Given $\bm{c}$, $\bm{j}$, and $\bm{m}$, Alice measures $n_{\rm rep}$ copies of system $A$ 
to obtain $\bm{a}$ and $\bm{t}$, which is represented by the POVM elements 
$\{\hat{T}^{(\bm{c},\bm{j},\bm{m})}_{\bm{a},\bm{t}}
\coloneqq \bigotimes_{r=1}^{n_{\rm rep}} 
\hat{T}^{(c_r,j_r,m_r)}_{a_r,t_r}\}$.

Let $p(\bm{c})$ be the probability of vector $\bm{c}$, and 
$p(\bm{c},\bm{a},\bm{b},\bm{j},\bm{t})$ be the joint probability for the five vectors. We then have
\begin{eqnarray}
p(\bm{c},\bm{a},\bm{b},\bm{j},\bm{t})
=\sum_{\bm{m}}p(\bm{c})~{\rm tr}\left((\hat{T}^{(\bm{c},\bm{j},\bm{m})}_{\bm{a},\bm{t}}
\otimes  \hat{D}_{\bm{j},\bm{b}}) (\hat{N}_{\bm{m}}\hat{\rho}_{AS}(\bm{c})\hat{N}_{\bm{m}})
\right).
\label{kakuritsu}
\end{eqnarray}
Let $g_{\bm{t},\bm{j}}(\bm{c})$ be a function for fixed $\bm{t}$ and $\bm{j}$ defined as 
$g_{\bm{t,j}}(\bm{c})=(\bar{c}_1,\bar{c}_2,..\bar{c}_{n_{\rm rep}})$ where 
$\bar{c}_r=c_r$ ($t_r$=1 or $j_r=0$) and $\bar{c}_r=0$ ($t_r$=0 and $j_r\neq 0$). 
From Eq.~(\ref{hitoshi}), for $t_r=0$ and $j_r \neq 0$ we have 
\begin{eqnarray}
{\rm tr}_A\left((\hat{T}_{a_r,0}^{(0,j_r,m_r)}\otimes \hat{\mathbbm{1}}_S)
(\hat{N}_{m_r}\hat{\sigma}_{AS}(0)\hat{N}_{m_r})\right) 
={\rm tr}_A\left((\hat{T}_{a_r,0}^{(1,j_r,m_r)}\otimes \hat{\mathbbm{1}}_S)
(\hat{N}_{m_r}\hat{\sigma}_{AS}(1)\hat{N}_{m_r})\right), \nonumber \\
\label{mister}
\end{eqnarray}
since $\hat{\sigma}_{AS}(c_r)= \ket{\Psi(c_r)}_{AS}\bra{\Psi(c_r)}$. 
 Thus, 
for $\bm{c}, \bm{c'}$ satisfying $g_{\bm{t,j}}(\bm{c})=g_{\bm{t,j}}(\bm{c'})=\bm{c}_{\rm const}$, 
we have 
\begin{equation}
{\rm tr}_A\left((\hat{T}^{(\bm{c},\bm{j},\bm{m})}_{\bm{a},\bm{t}}\otimes \hat{\mathbbm{1}}_S)
(\hat{N}_{\bm m}\hat{\rho}_{AS}(\bm{c})\hat{N}_{\bm m})\right)=
{\rm tr}_A\left((\hat{T}^{(\bm{c'},\bm{j},\bm{m})}_{\bm{a},\bm{t}}\otimes \hat{\mathbbm{1}}_S)
(\hat{N}_{\bm m}\hat{\rho}_{AS}(\bm{c'})\hat{N}_{\bm m})\right).
\end{equation}
Therefore, Eq.~(\ref{kakuritsu}) 
is written in the form 
$p(\bm{c},\bm{a},\bm{b},\bm{j},\bm{t})=p(\bm{c}) \beta (g_{\bm t,\bm j}(\bm{c}), \bm{a},\bm{b},\bm{j},\bm{t})$,  
which leads to, 
for a given value of $\bm{c}_{\rm const}$, we obtain 
\begin{eqnarray}
\frac{p(\bm{c},\bm{a},\bm{b},\bm{j},\bm{t})}
{\sum_{\bm{c'}:g_{\bm{t,j}}(\bm{c'})=\bm{c}_{\rm const}}p(\bm{c'},\bm{a},\bm{b},\bm{j},\bm{t})}
&=&\frac{p(\bm{c}) \beta (\bm{c}_{\rm const}, \bm{a},\bm{b},\bm{j},\bm{t})
 }{\sum_{\bm{c'}:g_{\bm{t,j}}(\bm{c'})=\bm{c}_{\rm const}}
 p(\bm{c'}) \beta (\bm{c}_{\rm const}, \bm{a},\bm{b},\bm{j},\bm{t}) }\nonumber \\
&=&\frac{p(\bm{c})}{\sum_{\bm{c'}:g_{\bm{t,j}}(\bm{c'})=\bm{c}_{\rm const}}p(\bm{c'})}
\label{ransom}
\end{eqnarray}
for $\bm{c}$ satisfying $g_{\bm{t,j}}(\bm{c})=\bm{c}_{\rm const}$. 
Eq.~(\ref{ransom}) shows that for the rounds with $t=0$ and $j\neq 0$, the probability of 
obtaining $c=0,1$ is $p_0$, $p_1$ and is independent 
of the value of $a,b,j$. 
Therefore, in the limit of $n_{\rm rep}\to \infty$, 
\begin{equation}
\frac{n(c=0, t=0, a\neq b, j\neq 0)}{n(c=1, t=0 , 
a \neq b, j\neq 0)}=\frac{p_0}{p_1}
\label{ransam2}
\end{equation}
holds, where $n(condition)$ denotes the
number of rounds satisfying the {\it condition} in the $n_{\rm rep}$ rounds. 
Finally, notice that Bob conducts check-basis measurement regardless of the 
value of $d$ 
in the alternative protocol, and hence $d$ is independent of the other variables. 
Therefore, we have 
\begin{equation}
\frac{n(c=d=0, t=0, a\neq b, j\neq 0)}{n(c=d=1, t=0,  
a \neq b, j\neq 0)}=\left(\frac{p_0}{p_1}\right)^2,
\label{ransam2}
\end{equation}
which corresponds to Eq.~(\ref{ransam}).

\section{Security proof for a general light source}
Here we show that our proof 
can be extended to the use of a general light source. 
Suppose that the laser in Fig.~\ref{setupfig} emits a train of $L$ pulses in a general mixed state $\hat{\sigma}_S$. 
We assume that every train from the laser is independent and has the same state $\hat{\sigma}_S$. 
We also assume that the subsequent phase modulation is ideal. 
The state after the phase modulation, which was given in Eq.~(\ref{prepa}) in the description of the actual protocol, 
is now given by 
\begin{equation}
\left(\bigotimes_{l=0}^{L-1}{\rm exp}\big(i \theta_l (a_l,c)\hat{m}_l\big)\right)~\hat{\sigma}_{S}~
\left(\bigotimes_{l'=0}^{L-1}{\rm exp}\big(-i \theta_{l'} (a_{l'},c)\hat{m}_{l'}\big)\right), \label{prepre}
\end{equation}
and the one after the randomization of the overall optical phase is    
\begin{equation}
\sum_m\hat{N}_m
\left(\bigotimes_{l=0}^{L-1}{\rm exp}\big(i \theta_l (a_l,c)\hat{m}_l\big)\right)~\hat{\sigma}_{S}~
\left(\bigotimes_{l'=0}^{L-1}{\rm exp}\big(-i \theta_{l'} (a_{l'},c)\hat{m}_{l'}\big)\right)
\hat{N}_m
\label{enuemu2}
\label{enuemu2}
\end{equation}
instead of Eq.~(\ref{enuemu}).

The security proof in Sec.~\ref{security proof} used the assumption of pure coherent states Eq.~(\ref{prepa}) in several occasions, which 
are listed as follows: \\
i) The state preparation in the alternative protocol 
[Eq.~(\ref{kprepa})], 
and its relation [Eq.~(\ref{puramai})] to the actual protocol.\\
ii) The parity correlation [Eq.~(\ref{korekore})] 
between the auxiliary qubits and the photon numbers in pulses.\\
iii) The derived properties [Eqs.~(\ref{ningyou}), (\ref{Psinituite}), (\ref{30}), (\ref{awaseru}), 
(\ref{hitoshi}) 
and (\ref{mister})] for proving that the sampling is unbiased as in Eq.~(\ref{ransam}).\\
iv)  The expressions [Eqs.~(\ref{tokkyu}) and (\ref{tagtag})] for the parameter $r_{\rm tag}$. \\
In what follows, we describe how each of the above arguments are rephrased in terms of the general state 
$\hat{\sigma}_S$.

i) In the alternative protocol, 
we assume that Alice prepares the following state on system $AS$,  
\begin{equation}
\hat{\sigma}_{AS}(c)\coloneqq \hat{R}(c)\hat{\sigma}_{S}\hat{R}(c),
\label{sigmaAS}
\end{equation}
instead of Eq.~(\ref{kprepa}). Here $\hat{R}(c)$ is defined by 
\begin{equation}
\hat{R}(c)\coloneqq \bigotimes_{l=0}^{L-1}
\Big[\frac{1}{\sqrt{2}}\Big(
\ket{+}_{A,l}{\rm exp}\big(i \frac{\pi}{2}lc\hat{m}_l\big)
+\ket{-}_{A,l}{\rm exp}\big(i (\pi+\frac{\pi}{2}lc)\hat{m}_l\big)\Big)
\Big].
\end{equation}
Then it is straightforward to confirm that 
\begin{eqnarray}
&&\left(\bigotimes_{l=0}^{L-1} {}_{A,l}\bra{\pm}\right)\hat{\sigma}_{AS}(c)
\left(\bigotimes_{l'=0}^{L-1}\ket{\pm}_{A,l'}\right)\nonumber \\
&=&
\frac{1}{2^L}
\left(\bigotimes_{l=0}^{L-1}{\rm exp}\big(i \theta_l (a_l,c)\hat{m}_l\big)\right)~\hat{\sigma}_{S}~
\left(\bigotimes_{l'=0}^{L-1}{\rm exp}\big(-i \theta_{l'} (a_{l'},c)\hat{m}_{l'}\big)\right),
\end{eqnarray}
where $\pm$ of the $l$-th qubit should be chosen according to the bit $a_l$. 
This is the general-state expression for Eq.~(\ref{puramai}), which leads to the equivalence of state preparation 
between the actual and the alternative protocol. 

ii) As $\hat{R}(c)$ is written in $Z$ basis as 
\begin{eqnarray}
\hat{R}(c)
&=&\bigotimes_{l=0}^{L-1} 
\Big[
\frac{1}{2}i^{~lc\hat{m}_l} 
\Big(
\ket{0}_{A,l}\big(\hat{\mathbbm{1}}_{S,l}+(-1)^{\hat{m}_l}\big)
+\ket{1}_{A,l}\big(\hat{\mathbbm{1}}_{S,l}-(-1)^{\hat{m}_l}\big)\Big)
\Big]
\nonumber
\\
&=&\bigotimes_{l=0}^{L-1} 
\Big[
i^{~lc\hat{m}_l} 
\Big(\ket{0}_{A,l}\sum_{m_l:{\rm even}}\hat{P}(\ket{m_l}_{S,l})
+\ket{1}_{A,l}\sum_{m_l:{\rm odd}}\hat{P}(\ket{m_l}_{S,l})\Big)\Big]\label{atodene}, 
\end{eqnarray}
we have 
\begin{equation}
\hat{\Upsilon}_{AS}\hat{R}(c)=\hat{R}(c), 
\label{kieru}
\end{equation}
which is a generalization of Eq.~(\ref{korekore}). 
It immediately implies that $\hat{\Upsilon}_{AS}\hat{\sigma}_{AS}(c)\hat{\Upsilon}_{AS}=\hat{\sigma}_{AS}(c)$,  
which indicates a property of state $\hat{\sigma}_{AS}$ that 
the measurement outcome on $Z$ basis $\{\ket{0}_{A,l},\ket{1}_{A,l}\}$ always coincides with the parity of photon 
number in the $l$-th pulse.

iii) 
 From (\ref{atodene}), we have 
\begin{equation}
\hat{R}(c)=\left(\bigotimes_{l=0}^{L-1}i^{~lc \hat{m}_l}\right)\hat{R}(0).
\label{Rnituite}
\end{equation}
Comparing Eqs.~(\ref{korekore}) and (\ref{Psinituite}) to 
Eqs.~(\ref{kieru}) and (\ref{Rnituite}), we see that the derived properties of 
Eqs.~(\ref{ningyou}), (\ref{30}), and (\ref{awaseru}) for $\ket{\Psi(c)}_{AS}$ 
should also hold for $\hat{R}(c)$. As a result, we obtain 
\begin{equation} 
{}_A\bra{\mathcal{A}_{a,\{z_l\}}^{(0,j)}}
\hat{N}_m\hat{R}(0)
=(-i)^{u(j)}{}_A\bra{\mathcal{A}_{a,\{z_l\}}^{(1,j)}}
\hat{N}_m\hat{R}(1)
~~{\rm for}~~\sum_{l\neq j}z_l=m
\label{gorgeous}
\end{equation}
as a generalization of Eq.~(\ref{hitoshi}). From Eq.~(\ref{gorgeous}), we have
\begin{eqnarray}
{}_A\bra{\mathcal{A}_{a,\{z_l\}}^{(0,j)}}\hat{N}_m \hat{\sigma}_{AS}(0) \hat{N}_m
\ket{\mathcal{A}_{a,\{z_l\}}^{(0,j)}}_A
=
{}_A\bra{\mathcal{A}_{a,\{z_l\}}^{(1,j)}} \hat{N}_m \hat{\sigma}_{AS}(1) \hat{N}_m
\ket{\mathcal{A}_{a,\{z_l\}}^{(1,j)}}_A \nonumber \\
~~~~~~~~~~~~~~~~~~~~~~~~~~~~~~~~~~~~~~~~~~~~~~~~~~~~~~~~~~~~~~~~~~~~~~~~~~~~~~{\rm for}~~\sum_{l\neq j}z_l=m,
\end{eqnarray}
which assures that Eq.~(\ref{mister}) is also true when $\hat{\sigma}_{AS}(c)$ is given by Eq.~(\ref{sigmaAS}). 
Hence, Eq.~(\ref{ransam}) holds.

iv) For the initial state given by Eq.~(\ref{sigmaAS}), the definition of the parameter $r_{\rm tag}$ of 
Eq.~(\ref{tokkyu}) 
is replaced by 
\begin{equation}
 r_{\rm tag}=1-\sum_m {\rm tr}\left((\hat{\Pi}^{(m)}_{A}\otimes \hat{N}_m) \hat{\sigma}_{AS}(0)\right).
\end{equation}
Together with Eqs.~(\ref{regardless}) and (\ref{kieru}), 
we have 
\begin{equation}
 r_{\rm tag}=1-\sum_m 
{\rm tr}\left((\hat{\mathbbm{1}}_A\otimes \hat{\Pi}^{(m)}_{S})
\hat{\sigma}_{AS}(0)\right)
=1-\sum_m {\rm tr}\left( \hat{\Pi}^{(m)}_{S}
\hat{\sigma}_{S}\right).
\label{rtag}
\end{equation}

\section{Calibration of light sources}
\begin{figure}[htbp]
  \begin{center}
  \raisebox{5mm}
  {\includegraphics[width=130mm]{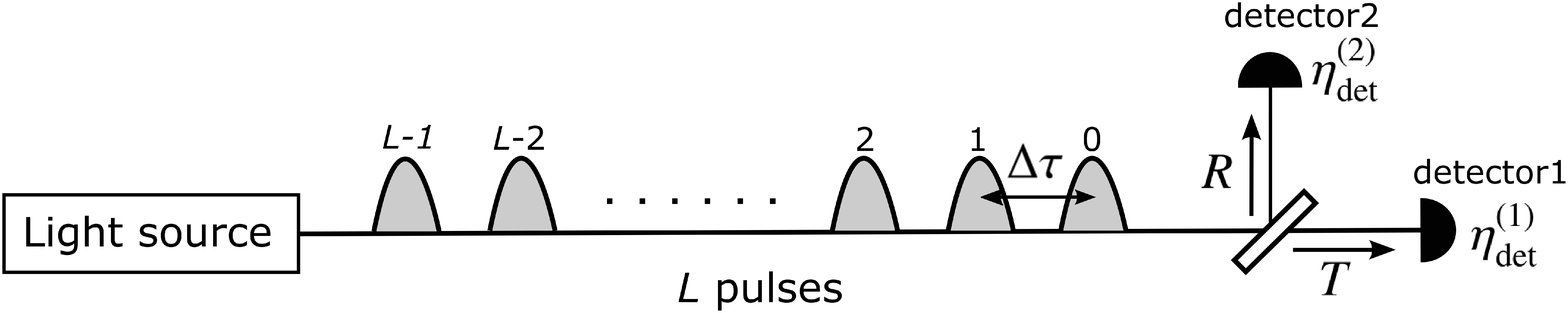}}
  \end{center}
  \begin{center}
  \caption{
  Off-line calibration setup to determine an upper bound on $r_{\rm tag}$ for a general light source, 
  when the dead time of detectors is shorter than pulse interval $\Delta \tau$. 
  $R$ and $T$ represent reflectance and 
  transmittance of the beam splitter, respectively. $\eta_{\rm det}^{(1)}$ and 
   $\eta_{\rm det}^{(2)}$ represent detection efficiencies of detector 1 and 2, respectively. 
    }
   \end{center}
  \label{calibration1}
\end{figure}
Here we discuss how we may determine an upper bound on 
the parameter $r_{\rm tag}$, which is given by  Eq.~(\ref{rtag}), from an off-line experiment on the light source. 
We use  
a beam splitter characterized by transmittance $T$ and reflectance $R$ and two threshold detectors 
with quantum efficiencies $\eta_{\rm det}^{(1)}$ and $\eta_{\rm det}^{(2)}$, as in Fig.~C1. 
No precise values of these parameters are needed, and we assume that there are known lower bounds  
$\eta_1\leq  T\eta_{\rm det}^{(1)}$ and $\eta_2\leq R\eta_{\rm det}^{(2)}$. 
For simplicity, we neglect the effect of dark countings of the detectors.
We assume that the dead time of the detectors are shorter than the pulse interval 
such that they are ready for every incident pulse. 
For an $L$ pulse train emitted from the source, 
we record the timings of detection at the two detectors, 
and define a double coincidence event to be the case when both detectors have reported 
detections within a pair of neighboring pulses.  

Since a state in the range of 
$\hat{\mathbbm{1}}-\sum_m \Pi_S^{(m)}$ contains at least two photons in a pair of neighboring pulses, 
such a state has a probability of resulting in a double coincidence event no smaller than 
$2\eta_1\eta_2$. 
Thus, if we repeat the measurement $n_{\rm test}$ times and find that double coincidence events 
have occurred 
$n_{\rm double}$ times, an upper bound on $r_{\rm tag}$ is given by
\begin{equation}
\bar{r}_{\rm tag}\coloneqq \frac{n_{\rm double}}{n_{\rm test}}\frac{1}{2\eta_1\eta_2}
\geq r_{\rm tag},
\label{timhor}
\end{equation}
in the asymptotic limit of large $n_{\rm test}$. 
Although the tightness of the upper bound  $\bar{r}_{\rm tag}$ varies depending on the state $\hat{\sigma}_S$ 
in general, we may show that it can be quite tight when the state is close to an ideal coherent state. 
Suppose that $\eta_1$ and $\eta_2$ are equal to the actual efficiencies, and each pulse 
is exactly in the coherent state with amplitude $\mu$. 
For every pulse, detector 1 and 2 independently report detection with probability 
$p_k^{(\rm click)}= 1-e^{-\eta_k \mu} \le \eta_k \mu~(k=1,2)$. 
Since there are $L+2(L-1)$ different combinations of timings leading to double coincidence, 
we have 
\begin{eqnarray} 
\frac{n_{\rm double}}{n_{\rm test}}
\leq(3L-2)p_1^{(\rm click)}p_2^{(\rm click)}
\leq  \eta_1\eta_2\mu^2 (3L-2),
\end{eqnarray}
which leads to 
\begin{equation}
\bar{r}_{\rm tag}\leq \frac{\mu^2 (3L-2)}{2}.
\end{equation}
On the other hand, direct calculation shows that, in the limit of $L\mu^2\to 0$, 
\begin{eqnarray}
r_{\rm tag}&=&
\mu^2\frac{3L-2}{2}+\mu^3\Big(\frac{-10L+12}{3}\Big)+\mu^4\Big(\frac{-9L^2+82L-120}{8}\Big)
+O(L^2 \mu^5+L^3 \mu^6)\nonumber \\
&=& \mu^2\frac{3L-2}{2}-\mu^3L\Big(\frac{9}{8}\mu L+\frac{10}{3}\Big)+O(L^2 \mu^5+L^3 \mu^6),
\label{3}
\end{eqnarray} 
which leads to   
\begin{equation}
\frac{\bar{r}_{\rm tag}-r_{\rm tag}}{r_{\rm tag}}\leq \mu\Big(\frac{3}{4}\mu L+\frac{20}{9}\Big)
+O(L \mu^3+L^2 \mu^4).
\label{relative}
\end{equation}
Hence, the bound $\bar{r}_{\rm tag}$ is a good approximation of $r_{\rm tag}$ for $\mu \ll L^{-1/2}$ .

\begin{figure}[htbp]
  \begin{center}
  \raisebox{5mm}
  {\includegraphics[width=150mm]{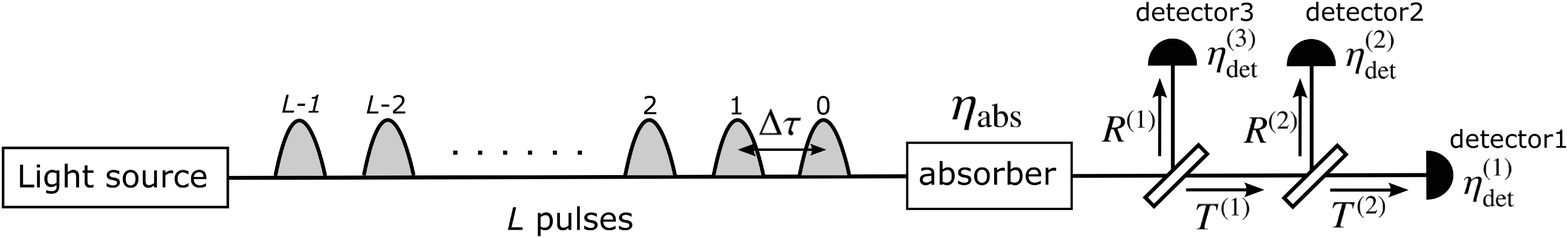}}
  \end{center}
  \begin{center}
  \caption{
    Off-line calibration setup to determine an upper bound on $r_{\rm tag}$ for a general light source, 
  when the dead time of detectors is longer than pulse interval $\Delta \tau$. 
  An optical linear absorber with transmittance $\eta_{\rm abs}$ is set in front of 
  beam splitters. $R^{(1)}$, $R^{(2)}$, 
  $T^{(1)}$ and $T^{(2)}$ represent reflectance and 
  transmittance of the two beam splitters. $\eta_{\rm det}^{(1)}$, 
   $\eta_{\rm det}^{(2)}$ and $\eta_{\rm det}^{(3)}$ represent detection efficiencies of 
   threshold detector 1, 2 and 3, respectively. 
   }
   \end{center}
  \label{calibration2}
\end{figure}

In a more practical case where the dead time ($\tau_{\rm dead}$) of the detectors is 
longer than the pulse interval 
$(\tau_{\rm dead} >\Delta \tau)$, 
there is a possibility that the presence of two photons is masked by an earlier detection of a third photon. 
In such a case, we may use a setup in Fig.~C2 with three detectors and a linear absorber with transmittance 
$\eta_{\rm abs}$.
Assume that we know lowerbounds, $\tilde{\eta}_{\rm abs}\leq \eta_{\rm abs}$, 
$\tilde{\eta}_1\leq T^{(1)}T^{(2)} \eta_{\rm det}^{(1)}$, 
$\tilde{\eta}_2\leq T^{(1)}R^{(2)} \eta_{\rm det}^{(2)} $ and 
$\tilde{\eta}_3\leq R^{(1)} \eta_{\rm det}^{(3)}$. 
Define a triple coincidence event to be the case when all three detectors has reported 
detections within the whole train of $L$ pulses. 
Let $q_3$ be the probability that the $L$ pulse train leaving the linear absorber contains three or more photons. 
If we repeat the measurement $n_{\rm test}$ times and 
triple coincidence events have occurred $n_{\rm triple}$ times, we have 
\begin{equation}
q_3\leq \frac{n_{\rm triple}}{n_{\rm test}}\frac{1}{6\tilde{\eta}_1\tilde{\eta}_2\tilde{\eta}_3}
\label{samui}
\end{equation}
in the limit of large $n_{\rm test}$. 
Suppose that one records the number
$n_{\rm double}^{\rm (obs)}$ 
 of double coincidence events in the same $n_{\rm test}$ runs, which is defined as the case 
 when detectors 1 and 2 have reported detections within a pair of neighboring pulses. 
 Since the effect of the dead time can be simulated with a fictitious detector with no dead time by 
 ignoring detection events that occurred when the real detector would have been dead, 
 we may consider the number
$n_{\rm double}^{\rm (true)}$ 
 of double coincidence events defined from these fictitious detectors. 
Since the 
two definitions of a double coincidence event differs only when three or more photons are incident on the two detectors,
we have 
\begin{equation}
\frac{n_{\rm double}^{(\rm true)}}{n_{\rm test}}\leq \frac{n_{\rm double}^{(\rm obs)}}{n_{\rm test}}+q_3
\label{q3dif}
\end{equation}
in the limit of large $n_{\rm test}$. 
On the other hand, as in Eq.~(\ref{timhor}), $n_{\rm double}^{(\rm true)}$ satisfies 
\begin{equation}
r_{\rm tag}\leq \bar{r}_{\rm tag}= \frac{n_{\rm double}^{(\rm true)}}{n_{\rm test}}\frac{1}{2\eta_1\eta_2}
\label{nemui}
\end{equation}
by taking $\eta_1=\tilde{\eta}_{\rm abs}\tilde \eta_1$ and 
$\eta_2=\tilde{\eta}_{\rm abs}\tilde \eta_2$. 
We thus obtain an upper bound from Eqs.~(\ref{samui})-(\ref{nemui}) as 
\begin{eqnarray}
r_{\rm tag} \leq \bar{r}_{\rm tag}^{~*} \coloneqq
\Big(\frac{n_{\rm double}^{(\rm obs)}}{{n_{\rm test}}}
+\frac{n_{\rm triple}}{n_{\rm test}}\frac{1}
{6\tilde{\eta}_1\tilde{\eta}_2\tilde{\eta}_3}\Big)
\frac{1}{2\tilde{\eta}_1 \tilde{\eta}_2 \tilde{\eta}_{\rm abs}^2}.
\label{hortons}
\end{eqnarray}

We show that $\bar{r}_{\rm tag}^{~*}$ also 
approximates $r_{\rm tag}$ well when the light source emits coherent pulses. 
Suppose that $\tilde{\eta}_1$, $\tilde{\eta}_2$, $\tilde{\eta}_3$ and 
$\tilde{\eta}_{\rm abs}$
are equal to the actual efficiencies. 
Since $n_{\rm double}^{(\rm obs)} \leq n_{\rm double}^{(\rm true)}$ holds,
 we have 
\begin{eqnarray}
\bar{r}_{\rm tag}^{~*}
\leq
\bar{r}_{\rm tag}+
\frac{n_{\rm triple}}{n_{\rm test}}\frac{1}{6\tilde{\eta}_1\tilde{\eta}_2\tilde{\eta}_3}
\frac{1}{2\tilde{\eta}_1 \tilde{\eta}_2 \tilde{\eta}_{\rm abs}^2}
.
\end{eqnarray}
From Eq.~(\ref{relative}), we have 
\begin{eqnarray}
\frac{\bar{r}_{\rm tag}^{~*}-r_{\rm tag}}{r_{\rm tag}}\leq 
\mu\Big(\frac{3}{4}\mu L+\frac{20}{9}\Big)
+\frac{n_{\rm triple}}{n_{\rm test}}\frac{1}{6\tilde{\eta}_1\tilde{\eta}_2\tilde{\eta}_3}
\frac{1}{2\tilde{\eta}_1 \tilde{\eta}_2 \tilde{\eta}_{\rm abs}^2} \frac{1}{r_{\rm tag}}+O(L \mu^3+L^2 \mu^4)
\nonumber \\
\end{eqnarray}
for $L \mu^2 \to 0$. 
Since $L$ pulses incident on detector $k$ lead to one or more detections at probability 
$p_k^{(\rm click)}=1-e^{-\tilde{\eta}_k\tilde{\eta}_{\rm abs}L\mu}\leq \tilde{\eta}_k \tilde{\eta}_{\rm abs}L\mu$, we have 
\begin{equation}
\frac{n_{\rm triple}}{n_{\rm test}}
\leq 
\tilde{\eta}_1\tilde{\eta}_2\tilde{\eta}_3(\tilde{\eta}_{\rm abs}L\mu)^3.
\end{equation}
Thus, we obtain 
\begin{eqnarray}
\frac{\bar{r}_{\rm tag}^{~*}-r_{\rm tag}}{r_{\rm tag}}&\leq& 
\mu\Big(\frac{3}{4}\mu L+\frac{20}{9}\Big)
+\frac{(\tilde{\eta}_{\rm abs}L\mu)^3}{12\tilde{\eta}_1 \tilde{\eta}_2 \tilde{\eta}_{\rm abs}^2}\frac{1}{r_{\rm tag}}
+O(L \mu^3+L^2 \mu^4)\nonumber \\
&=&\mu\Big(\frac{3}{4}\mu L+\frac{20}{9}\Big)+\mu \frac{\tilde{\eta}_{\rm abs} L^2}{18\tilde{\eta}_1 \tilde{\eta}_2 }
+O(L \mu^3+L^2 \mu^4). \label{bomb}
\end{eqnarray}
Therefore, $\bar{r}_{\rm tag}^{~*}$ becomes a good approximation of $r_{\rm tag}$ when $\mu \ll L^{-1/2}$ 
and the absorber is chosen to satisfy $\tilde{\eta}_{\rm abs} \mu  \ll L^{-2}$.

\section*{References}
\bibliographystyle{iopart-num}
\bibliography{sasakibib,NJPyoubib}

\end{document}